\documentclass[a4paper, amsfonts, amssymb, amsmath, reprint, showkeys, nofootinbib, twoside, superscriptaddress]{revtex4-2}
\usepackage[english]{babel}
\usepackage[utf8]{inputenc}

\usepackage[pdftex, pdftitle={Article}, pdfauthor={Author}]{hyperref} % For hyperlinks in the PDF

\usepackage{graphicx}% Include figure files
\usepackage{dcolumn}% Align table columns on decimal point
\usepackage{bm}% bold math
\usepackage{siunitx}
\usepackage{pstricks}
\usepackage{comment}
\usepackage{booktabs}
\usepackage{setspace}
\usepackage{ragged2e} % For text justification
\usepackage{upgreek}
\usepackage{braket}
\usepackage{siunitx}
\usepackage{xcolor}  
\usepackage{gensymb}
\DeclareSIUnit\bar{bar}
\DeclareSIUnit\cps{cps}

\newcommand{\eup}{\mathrm{e}}
\usepackage{textgreek}

\begin{document}

\DeclareSIUnit\bar{bar}
\DeclareSIUnit\cps{cps}
%\DeclareMathOperator{\sinc}{sinc}
%\usepackage{textgreek}

%%%%%%%%%%%%%%%%%%%%%%%%%%%%%%%%%%%%%%%%%%%%%%%%%%%%%%%%%%%%

\title{An optical frequency shifter based on continuous-wave pump fields} 

\author{Anica Hamer}
\affiliation{Physikalisches Institut, Rheinische Friedrich-Wilhelms-Universität Bonn, Bonn, Germany}
\author{Frank Vewinger}
\affiliation{Institut für Angewandte Physik, Rheinische Friedrich-Wilhelms-Universität Bonn, Bonn, Germany}
\author{Michael H.~Frosz}
\affiliation{Max-Planck-Institut für die Physik des Lichts, Erlangen, Germany}
\author{Simon Stellmer}
 \email{stellmer@uni-bonn.de}
\affiliation{Physikalisches Institut, Rheinische Friedrich-Wilhelms-Universität Bonn, Bonn, Germany}
\date{\today}% It is always \today, today,
             %  but any date may be explicitly specified

\begin{abstract}
Practical implementations of quantum information networks require frequency conversion of individual photons. Approaches based on a molecular gas as the nonlinear medium cover a wide range of the optical spectrum and promise high efficiency at negligible background. We present polarization-preserving frequency conversion in a hydrogen-loaded hollow core fiber using continuous-wave pump fields. We demonstrate conversion efficiency at the level of a few per mille, discuss various limitations and loss mechanisms, and present a route to increase conversion efficiency to near unity.
\end{abstract}

%\keywords{Suggested keywords}%Use showkeys class option if keyword
                              %display desired
\maketitle

%\tableofcontents

\section{Introduction}

Quantum computing and quantum communication will be based on hybrid architectures, at least partially in the optical domain. To facilitate interaction among different components, the information transfer via single photons needs to preserve the quantum state \cite{Kimble2008,Ruetz2017,Zaske2012,Krutyanskiy2017,Fisher2021,Bell2017,Zhou2014,Ikuta2011,DeGreve2012,Tyumenev2022,Winzer2018,Deutsch2023}. The process is called quantum frequency conversion and relies on the nonlinearity of materials. The most well-known experiments are either using the $\chi^{(2)}$ nonlinearity of crystals \cite{Ruetz2017, Zaske2012, Fisher2021, Bell2017, Zhou2014, Ikuta2011,DeGreve2012} and the $\chi^{(3)}$ nonlinearity of atomic and molecular gases \cite{Eramo1994,Tamaki1998,Babushkin2008,Cassataro2017,Tyumenev2022,Rowland2024}.

Crystals are known for their compactness combined with high conversion efficiency, but unfortunately also for their incoherent background and narrow spectral bandwidth. The $\chi^{(3)}$ nonlinearity of molecular gases, on the other hand, can take advantage of its high acceptance bandwidth and has also demonstrated very high conversion efficiencies when using pulsed lasers \cite{Tyumenev2022} or when operated close to atomic resonances \cite{Rowland2024}. In particular, coherent Stokes and anti-Stokes Raman scattering (CSRS/CARS) processes show great potential, as they can utilize virtual energy levels, as demonstrated in Refs.~\cite{Boyd2008,Shipp2017,Rigneault2018,Groessle2020,Tian2023, Gonzalez2025}.
 
Here, we present frequency conversion from the near-infrared (NIR) to the telecom S-band (\SIrange{1460}{1530}{\nano\meter}) by using the CSRS process in a hydrogen-filled anti-resonant reflecting hollow-core fiber (HCF) with continuous-wave lasers. The here used signal wavelength (\SI{1474}{\nano\meter})\cite{Javadi2015}, along with other wavelengths from previous experiments \cite{Aghababaei2023, Hamer2024a}, is of interest as it serves as a test wavelength for quantum dots, which facilitate interactions between different quantum systems. 

In previous works \cite{Hamer2024a, Hamer2025}, we showed polarisation preservation, the importance of optimized incoupling and homogenous pressure as well as the dependence of the intensity of the pump fields. The efficiency of the system reached a peak of \SI{5.2e-5}{\percent \per \watt^2} for a \SI{6}{\centi\meter} HCF.    
We build on previously gained insights showing now the dependence of the conversion efficiency on the fiber length leading to an efficiency of \SI{0.266}{\percent}. Furthermore, we comment on the HCF-bend radius and discuss the potential challenges related to attenuation and undesired Raman scattering.

%%%%%%%%%%%%%%%%%%%%%%%%%%%%%%%%%%%%%%%%%%%%%%%%%%%%%%%%%%%%%%%%%%%%%%%%%%%%%%%%%%%%

\section{Experimental setup}
We implement a frequency conversion from \SI{914}{\nano\meter} to \SI{1474}{\nano\meter}. 

Here, this work employs the same system as in our previous work \cite{Hamer2025} with respect to the laser system and the main experimental system. We are here solely outlining the key parts of the experiment, please find details in \cite{Hamer2024a, Hamer2025}.
\begin{figure}[htpb]
	\centering	  \includegraphics[width=\linewidth]{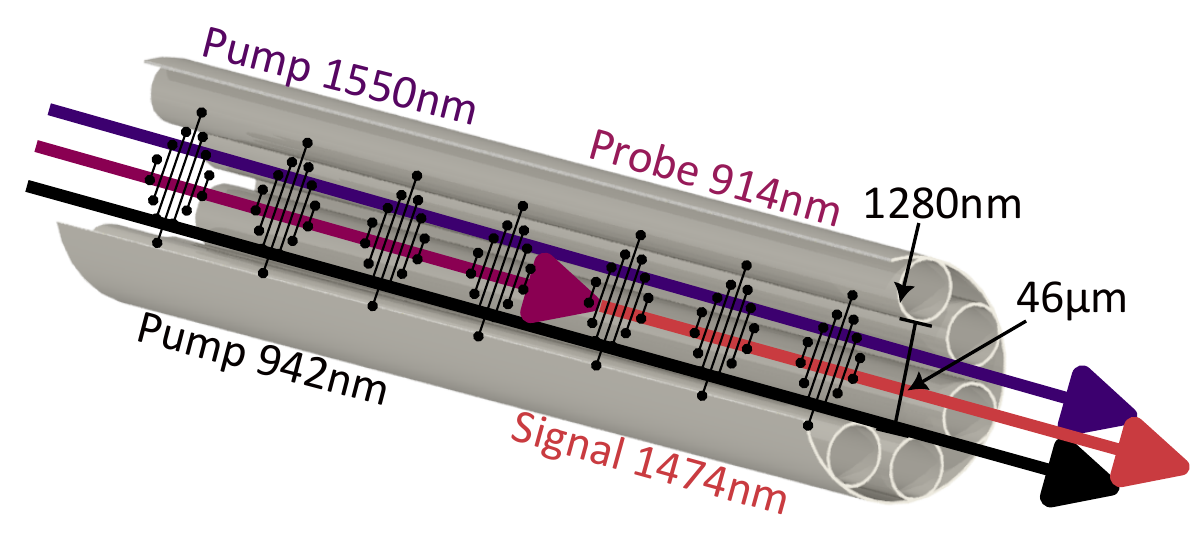}	
	\caption{Schematic overview of the fiber and the four light fields.}
	\label{fig:overview}
\end{figure}

The so-called pump lasers used in this work are essential for the CSRS process. They consist of two fields at \SI{1550}{\nano\meter} and \SI{942}{\nano\meter}, matching which their beat the vibrational transition at \SI{125}{\tera\hertz} within the $Q_1(1)$ branch of molecular hydrogen. The resulting polarization can be probed with \SI{914}{\nano\meter}, generating signal light at \SI{1474}{\nano\meter}, see Fig.~\ref{fig:overview}. The 942-nm pump field is generated by a VECSEL, the 1550-nm pump field is generated by a diode laser amplified by a fiber amplifier, and the 914-nm probe field is derived from an attenuated diode laser.

\begin{figure}[htpb]
	\centering
	  \includegraphics[width=\linewidth]{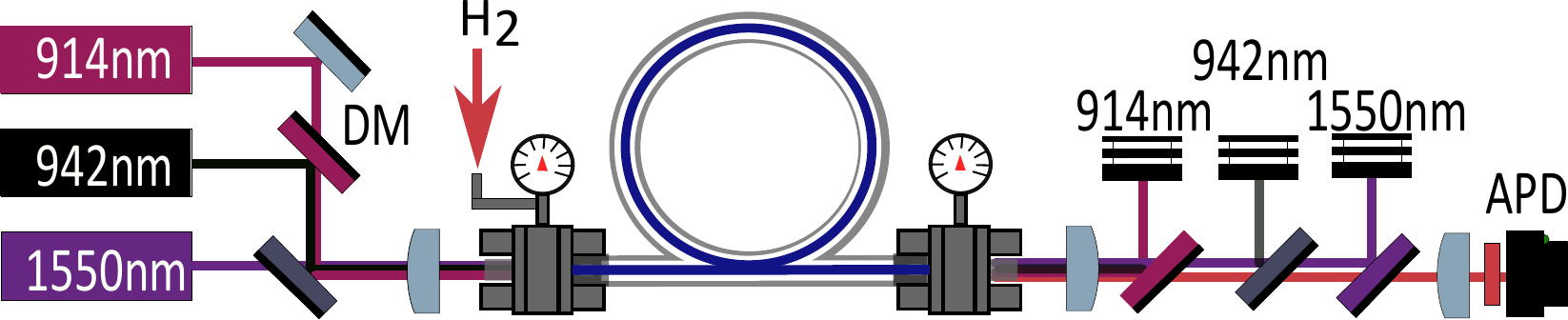}	
	\caption{Schematic depiction of the fiber (blue) inside a high pressure pipe (grey), connected to the entrance chambers. The two pump fields and the probe field are injected before, and dumped after, the fiber; the signal photons at 1474\,nm are detected on an avalanche photo detector (APD).}
	\label{fig:pipe}
\end{figure}

The fiber is placed inside a high pressure stainless steel pipe with an outer diameter of \SI{3}{\milli\meter} and an inner diameter of \SI{1.6}{\milli\meter}. On each end, the pipe is sealed to an entrance chamber, which is equipped with an AR-coated entrance window and a gas inlet; see Fig.~\ref{fig:pipe}. The fiber tips are held in place by a centering piece and a ferrule. No loss in gas pressure is observed in this system.

We employ a commercially available ARR-HCF fiber designed for \SI{1550}{\nano\meter}, similar to our previous work Ref.~\cite{Hamer2025}. The core diameter is about \SI{46}{\micro\meter}, and the seven capillaries surrounding the core have a thickness of around \SI{1280}{\nano\meter} (Fig.~\ref{fig:overview}). The latter value is the result of the comparison of experimental data on the optimal pressure for conversion and the theoretical calculations using the model given by Zeisberger-Hartung-Schmidt \cite{Zeisberger2018}. The maximal fiber length of \SI{1.85}{\meter} is reduced in the course of the experiment.

We employ one avalanche photodiode (APD) operated at \SI{10}{\percent} detection efficiency with a dead time of \SI{1}{\micro\second}, resulting in a dark count rate of approximately \SI{270}{\cps}. 

%%%%%%%%%%%%%%%%%%%%%%%%%%%%%%%%%%%%%%%%%%%%%%%%%%%%%%%%%%%%%%%%%%%%%%%%%%%%%%%%%%%%

\section{Frequency Conversion}\label{sec:freq}
We are measuring the efficiency $\eta$ as the photon rate of the converted \SI{1474}{\nano\meter} light divided by the photon rate of the incoming \SI{914}{\nano\meter} light. The efficiency scales as

\begin{equation}\label{eq:I}
    \eta \propto {| \chi^{(3)}(\omega) |}^2 L^2 {\mathrm{sinc}^2}\left(\frac{\Delta \beta (p) \cdot L }{2}\right) I_\text{pump1} I_\text{pump2},
\end{equation}
where $\chi^{(3)}(\omega)$ denotes the third-order nonlinear susceptibility and $L$ the interaction length \cite{Rigneault2018}. 
Furthermore, the phase matching condition $\Delta \beta = -\beta_\text{pump1} +\beta_\text{pump2}+\beta_\text{probe}-\beta_\text{signal}$ relies on the propagation constant $\beta = 2 \pi \nu/c \cdot  n_{\text{eff}}(p,\lambda)$, where the effective refractive index $n_{\text{eff}}$ is obtained from the Zeisberger model \cite{Zeisberger2018} and depends on the pressure $p$ and the wavelength $\lambda$.

\subsection{Efficiency}
\begin{figure}[htpb]
	\centering	\includegraphics[width=\linewidth]{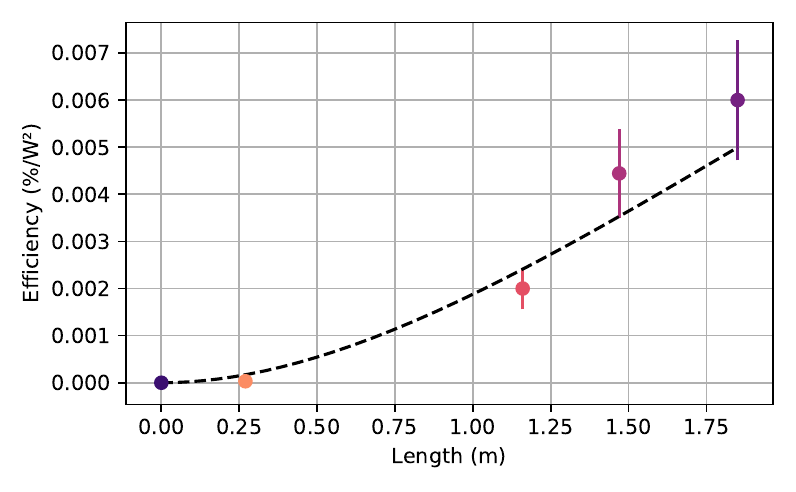}	
	\caption{Conversion efficiency in dependence of fiber length, using \SI{3}{\watt} of each pump field and otherwise identical parameters. A fit according to Eq.~\ref{eq:I}, corrected for absorption losses, is applied to the data (dashed line) and shows the quadratic increase with fiber length.}
	\label{fig:eff}
\end{figure}

While in previous experiments we demonstrated the quadratic dependence of the efficiency (Eq.~\ref{eq:I}) on the power of the pump fields for a fixed fiber length, we now show the expected quadratic dependence on the interaction length, i.e., the fiber length (Fig.~\ref{fig:eff}). 

To measure the values, we perform a cut-back of the fiber starting with a length of $L_1 = \SI{1,85 \pm 0,02}{\meter}$ followed by $L_2 =  \SI{1,47 \pm 0,02}{\meter}$, $L_3= \SI{1,16\pm0,02}{\meter}$ and $L_4 = \SI{0,27 \pm 0,02}{\meter}$.

For a maximum length of $L = \SI{1.85}{\meter}$, we obtain a fitted internal efficiency of \SI{0.006 \pm 0.0013}{\percent\per \watt^2} with a scaling of $\SI{0.0044 \pm 0.0006}{\percent \per (\watt^2 \meter^2)}$. Here, we have already included the transmission losses in the fit using the results described in the following subsection ~\ref{subsec:transmission}. This data was obtained with \SI{3}{\watt} of each pump laser. For the maximum power in the pump fields currently available to the experiment, \SI{3.87}{\watt} at \SI{942}{\nano\meter} and \SI{12.6}{\watt} at \SI{1550}{\nano\meter}, we obtain an internal efficiency of $\eta_\text{max} = \SI{0.27}{\percent}$.
We conclude that the length of the fiber could be increased further to improve the efficiency. 

\subsection{Losses}
An increase in fiber length will also increase the total, eventually defining an optimum length. Here, we study three loss mechanisms in detail: leakage loss \cite{Tyumenev2022, Zeisberger2018}, bend loss \cite{Frosz2016, Carter2017, NumkamFokoua2023}, and absorption loss due to hydrogen absorption \cite{Bahari2022}. Imperfection losses and material absorption have been described in Ref.~\cite{Yu2016} and will be ignored throughout this work.

\subsubsection{Transmission loss}\label{subsec:transmission}
Coupling between the core modes and capillary wall modes contributes to the transmission losses \cite{Tyumenev2022,Zeisberger2018}. They occur as broad resonances in the absorption spectrum, where the position of the resonances depends critically on the wall thickness of the capillaries. For the fiber used here, the wall thickness is about \SI{1.28}{\micro\meter}.

\begin{figure}[htpb]
	\centering    
    \includegraphics[width=\linewidth]{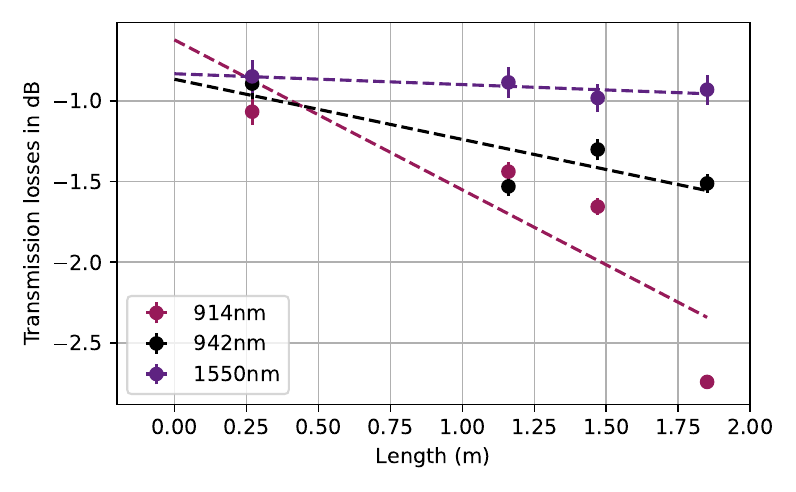}	
	\caption{Transmission losses for the three incoming light fields, depending on the length of the fiber. Linear curves are fitted to each data set (dashed lines). The intercept accounts for incoupling losses.}
	\label{fig:transmission}
\end{figure}

Here, we determine the loss parameter $\alpha$ through cut-back measurements, where we record the transmission of the Gaussian-like HE$_{11}$ fiber mode for various lengths, see Fig.~\ref{fig:transmission}.

Transmission loss is hardly noticeable for \SI{1550}{\nano\meter} with a loss parameter of $\alpha_{\text{1550\,nm}}=\SI{ 0.07 \pm 0.04}{\decibel \per \meter}$, which is expected since the fiber is designed for this wavelength (lower as \SI{0.035}{\decibel \per \meter} given by the manufacturer). However, there is a resonance centered around \SI{1330}{\nano\meter}, and the other two wavelengths are located only in the third transmission window, where losses are significant \cite{Zuba2023}. Here, we measure $\alpha_{\text{942\,nm}}=\SI{ 0.37 \pm 0.16}{\decibel \per \meter}$ for the second pump wavelength and $\alpha_{\text{914\,nm}}=\SI{0.93 \pm 0.37}{\decibel \per \meter}$ for the probe photons. The position of the resonances can be shifted through proper choice of the capillary wall thickness, and selection of a different fiber is expected to significantly reduce losses at the relevant wavelengths.

\subsubsection{Bend losses}
Another aspect of the transmission losses (Fig.~\ref{fig:transmission}) are bend losses. Due to the length of the fiber, the fiber needed to be wound (Fig.~\ref{fig:pipe}) to fit on the optical table. Depending on the bend radius, the core mode can couple to cladding capillary modes, leading to high losses \cite{Frosz2016, Carter2017, NumkamFokoua2023}.   

We measured the optimal pressure $p_\text{opt}$ relevant for the effective refractive index $n_\text{eff}$ and thus the phase matching condition $\Delta \beta$ (Eq.~\ref{eq:I}) in dependence of the bend radius.

\begin{figure}[htpb]
	\centering	\includegraphics[width=\linewidth]{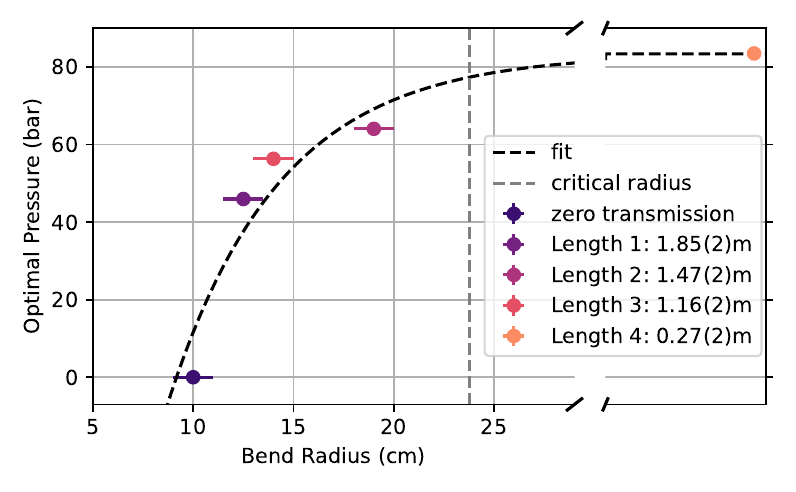}
	\caption{Optimal pressure for the CSRS process in dependence of the bend radius. Measurements were taken with different fiber lengths.
    To guide the eye, a simple saturation curve (Eq.~\ref{eq:popt}) is fitted to the data (black dashed line), and critical bend radius (vertical dashed line) is added according to Eq.~\ref{eq:bend}. See the text for more details.}
	\label{fig:bend}
\end{figure}

The data is shown in Fig.~\ref{fig:bend}. For a bend radius $r_0$ smaller than \SI{10 \pm 1}{\centi\meter}, no coupling to the  LP$_{01}$ mode of probe wavelength is possible and only the LP$_{11}$ mode can be addressed.

Furthermore, the fourth fiber of \SI{0.27 \pm 0.2}{\meter} length was not wound, but for a comparison the data point was shifted to a value magnitudes higher than the critical bend radius of \SI{24}{\centi\meter}, thus beyond the threshold where resonant bend losses occur.  
The critical bend radius $R_{l,m}$ for the LP$_{l,m}$ (here LP$_{0,1}$) mode is calculated with \cite{Carter2017, Frosz2016}   
\begin{eqnarray}\label{eq:bend}
    R_{l,m} = d / \left(\sqrt{\frac{n_{l,m}(\lambda)_\text{core}}{n_{l,m}(\lambda)_\text{cladding}}} -1 \right),
\end{eqnarray}
where $d$ is the distance between the center of the core and the capillary projected onto the plane of the bend, assuming that the bend plane and the fiber structure are aligned to obtain the largest value of the critical bend radius. The effective indices of the core and the cladding $n_{l,m}(\lambda)$ \cite{Martcatili1964} can be calculated via 
\begin{eqnarray}
    n_{l,m}(\lambda) = 1-\frac{1}{2}\left(\frac{j_{l,m}\lambda}{2\pi r}\right)^2.
\end{eqnarray}
The parameters are wavelength $\lambda$, $r$ as the effective radius of the hollow core and $j_{l,m}$ is the $m$th zero for the Bessel function of the first kind $J_l$. 

As a guide to the eye the complete data set is fitted to a  saturation curve 
\begin{eqnarray}\label{eq:popt}
    p_\text{opt} =  p_\text{max} \cdot (1-\eup^{{-b \cdot (r-r_0)}})
\end{eqnarray}
where $p_\text{max}$ is the optimal pressure reached with a straight fiber, $b$ is the saturation parameter, $r$ is the bend radius and $r_0$ is the shift due to the complete transmission loss of the LP$_{01}$ mode. 

As seen in Fig.~\ref{fig:bend}, there is a general trend towards higher pressures, reaching up to \SI{83 \pm 2}{\bar} with increasing bend radius. The deviations of the data points from the curve may indicate that other parameters, such as additional strain or mechanical stress on the fiber, possibly induced by unintended twisting, may play a role as well.
 
\subsubsection{Parasitic Raman processes in hydrogen}
Not only is efficiency an important factor for the usability of frequency conversion, but also a low background. While frequency conversion in crystals often struggles with undesired Raman scattering, it is minimized in hollow-core fibers due to the minimal overlap between the cladding and the core mode \cite{Tyumenev2022}.

\begin{figure}[htpb]
	\centering
	\includegraphics[width=\linewidth]{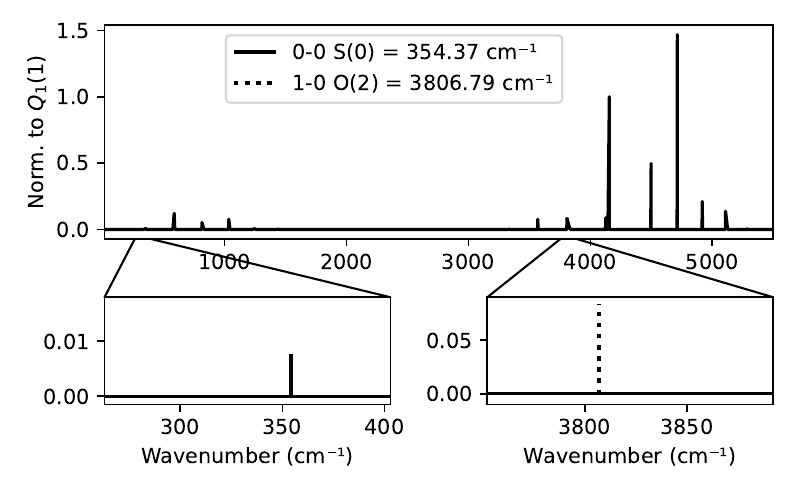}	
    \caption{Strength of the H$_2$ absorption lines in the wavenumber range of \SIrange{0}{5500}{\per\centi\meter}, normalized to the strength of the Q$_1(1)$ transition.} 
	\label{fig:background}
\end{figure}

Another type of background could originate from parasitic Raman processes in hydrogen gas itself. As the bandpass filter utilized to separate the light converted to the telecom band has a bandwidth of \SI{25}{\nano\meter}, we have investigated if parasitic processes are present. For this, we have examined if near-resonant Raman processes are possible, which lead to Stokes or anti-Stokes radiation near \SI{1474}{\nano\meter}. We have focused only on strong transitions, which are maximal two orders of magnitude smaller than the normally used $Q_1(1)$ transition, which are expected to dominate the background signal. An overview of the absorption lines is obtained from Refs.~\cite{hitran, Black1987} and shown in Fig.~\ref{fig:background}.

Indeed, we are detecting a background that is significantly stronger than in previous setups \cite{Hamer2025} and arises  for each pump beam independently and only in presence of hydrogen. We conclude that these Raman effects are based on the following two hydrogen lines (Fig.~\ref{fig:background} zoom-in), whose notation of the vibrational Raman spectrum lines can be read as follows: the vibrational quantum numbers of the upper and lower levels are given as $v'-v$, followed by the rotational branch label of Q ($\Delta J = 0$), O ($\Delta J = +2$), or S ($\Delta J = -2$) where in parentheses the lower-level rotational quantum number $J$ is shown.

Firstly, photons from the \SI{942}{\nano\meter} light field can be Raman-shifted to \SI{1468.6}{\nano\meter} via the $v'=1,\, J'=0 \rightarrow v=0,\, J=2$ (1-0 O(2)) transition at $3806\,\textrm{cm}^{-1}$. Such anti-Stokes photons pass all spectral filters and generate a background of about \SI{7e5}{\cps} for a pump field of \SI{4}{\watt}.

Secondly, the 1550\,nm pump field de-excites molecules from the $v'=0,\, J'=2$ state at an energy of $354\,\textrm{cm}^{-1}$ into the the $v=0,\, J=0$ ground state via conversion to a photon at \SI{1469.3}{\nano\meter} using the pure rotational 0-0 S(0) line. We observe a background rate of \SI{5.5e5}{\cps} for a pump field of \SI{12.5}{\watt}.

\begin{figure}[htpb]
	\centering
	\includegraphics[width=\linewidth]{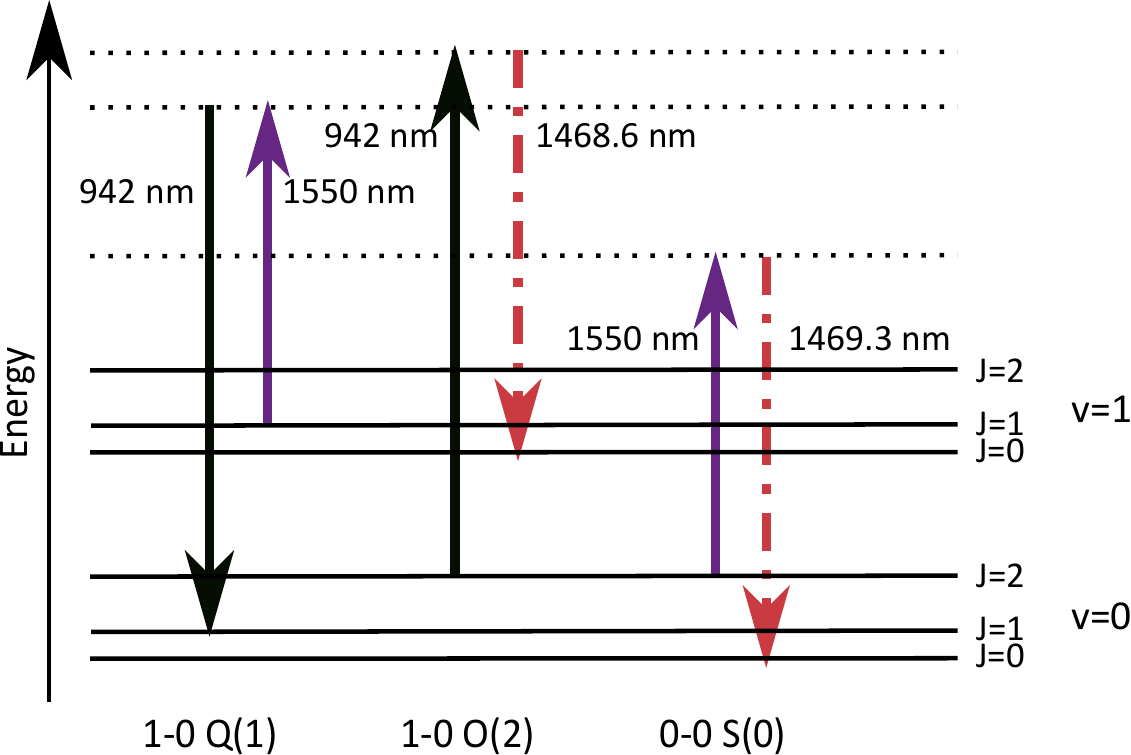}	
        \caption{Schemtic overview of the here used Raman line Q$_1(1)$ (1-0 Q(1)) and the parasitic Raman lines 1-0 O(2) and 0-0 S(0) \cite{hitran, Black1987}.}
	\label{fig:background2}
\end{figure}

A schematic comparison to the main Q$_1(1)$ (1-0 Q(1)) line is provided in Fig.~\ref{fig:background2}. While the overall absorption of the pump field is negligible, the signal-to-background ratio degrades significantly. To remove this background, very narrow bandpass filters or filter cavities can be used. Depending on the application, it might also be possible to shift the probe wavelength, and thus the signal wavelength, to a region where the pump fields do match a conversion line.

No Raman-induced background originating from the probe light was detected.

%%%%%%%%%%%%%%%%%%%%%%%%%%%%%%%%%%%%%%%%%%%%%%%%%%%%%%%%%%%%%%%%%%%%%%%%%%%%%%%

\section{Conclusion}

In summary, we demonstrated the scaling of conversion efficiency with fiber length. We achieved a maximum internal efficiency of 

\SI{0.0044(6)}{\percent\per\watt^2\per\meter^2}. For the 1.85-meter fiber used here and available pump powers of \SI{8}{\watt} (at \SI{942}{\nano\meter}) and \SI{15}{\watt} (at \SI{1550}{\nano\meter}), this would yield a theoretical maximum efficiency of \SI{1.9}{\percent}. In the experiment, losses from otical components and propagation losses within the fiber reduce the efficiency to \SI{0.27}{\percent}.

Increasing the efficiency further could be achieved by an increase in fiber length and increased pump power. In earlier work, we have worked with a fiber of smaller transmission losses at wavelengths away from hydrogen absorption lines \cite{Hamer2025}. Assuming the quadratic scaling with fiber length, incoupling efficiency of \SI{83}{\percent} and an attenuation of 15.9\,dB/km for all wavelengths, the optimum fiber length is about \SI{21}{\meter}, and, for pump powers of \SI{8}{\watt} and \SI{15}{\watt}, we arrive at a conversion efficiency of \SI{70}{\percent}. This value is on par with the performance achieved in pulsed operation \cite{Tyumenev2022}.

Such efficiency could also be achieved by higher pump powers, which would allow for a shorter fiber and thus reduce transmission losses. A substantially higher efficiency could also be achieved through use of a different molecular gas with asymmetric vibrational modes and potentially stronger third-order nonlinearities \cite{Ward1979}.

An additional contribution to losses is bending. It not only leads to the breakdown of transmission below a certain critical radius \cite{NumkamFokoua2023}, but also causes the optimal pressure to be reached at lower values. This is unfavorable for the conversion process, as the susceptibility scales quadratically with the number of molecules and thus negatively affects the efficiency \cite{Rigneault2018}. We find that for bending radii above 30\,cm, this effect becomes negligible.

The wavelength of the pump fields can be chosen freely, provided that their difference matches the $Q_1(1)$ resonance. Care should be taken to avoid pump fields interacting with hydrogen and creating photons in the signal region via Raman scattering.

%\begin{backmatter}
%\bmsection
{Funding - }
We acknowledge funding by Deutsche Forschungsgemeinschaft DFG through grant INST 217/978-1 FUGG and through the Cluster of Excellence ML4Q (EXC 2004/1 – 390534769), as well as funding by BMBF through the QuantERA project QuantumGuide.

%\bmsection
{Acknowledgments - }
We thank all members of the Cluster of Excellence ML4Q and the QuantumGuide collaboration, especially Thorsten Peters, as well as Philipp Hänisch, for stimulating discussions. 

%\bmsection
{Disclosures - }
The authors declare no conflicts of interest.

%\bmsection
{Data Availability Statement - }
Data underlying the results presented in this paper are not publicly available at this time but may be obtained from the authors upon reasonable request.

%\end{backmatter}

\bibliography{bib}
%\bibliographyfullrefs{bib}

\end{document}